\documentstyle[aps,preprint,eqsecnum]{revtex}

\input{epsf.sty}
\tightenlines

\begin{document}

\title{New algorithm and results for the 
three-dimensional random field Ising Model}
\author{J. Machta}
\address{Department of Physics and Astronomy,
University of Massachusetts,
Amherst, MA 01003--3720}
\author{M. E. J. Newman}
\address{Santa Fe Institute, 1399 Hyde Park Road, Santa Fe, NM 87501}
\author{L. B. Chayes}
\address{Department of Mathematics,
University of California,
Los Angeles, CA 90095--1555}

\date{\today}

\maketitle

\begin{abstract}
The random field Ising model with Gaussian disorder is
studied using a new Monte Carlo algorithm.  
The algorithm combines the advantanges of the replica exchange method 
and the two-replica cluster method and is much more efficient than the
Metropolis algorithm for some disorder realizations. Three-dimensional
sytems of size $24^3$ are studied.  Each realization of disorder is simulated
at a value of temperature and uniform field that is adjusted to the phase
transition region for that disorder realization.  Energy and
magnetization distributions show large variations from
one realization of disorder to another.  For some realizations
of disorder there are three well separated peaks in the
magnetization distribution and two well separated peaks in the
energy distribution suggesting a first-order transition.

\end{abstract} 


\section{Introduction}

Despite twenty five years of experimental and theoretical effort, phase
transitions in systems with quenched random fields are still poorly
understood.  The simplest theoretical model is the random field Ising model
(RFIM).  The RFIM phase transition is believed to be in the same
universality class as the phase transitions
 in diluted antiferromagnets in
a uniform field and fluids in porous media.  The three dimensional RFIM is
known~\cite{ImMa75,Imbrie84,BrKu87} to have an ordered phase at
sufficiently low temperature and for weak random fields.  As the
temperature or the strength of the randomness is increased, there is a
transition to a disordered phase.  The nature of this transition is not
well understood.

In this paper we describe a new replica exchange algorithm for simulating
the RFIM and present numerical results for systems of $L\times L\times L$
spins with $L$ up to~24.  We show that the qualitative features of the
transition differ strongly from realization to realization of disorder.
Our results can be interpreted as suggesting that the RFIM transition is
first-order or that a modified version of the droplet picture holds.

The RFIM is described by the energy
\begin{equation}
\label{eq:define}
-{\cal H}/k_{B}T = \beta \sum_{\langle ij \rangle} S_{i}S_{j} + 
\Delta \sum_{i} h_{i}S_{
i} + H \sum_{i} S_{i}
\end{equation}
where the spin variables $S_{i}=\pm 1$ reside on a lattice,
$\beta=1/k_{B}T$ is the inverse temperature, the first sum is over nearest
neighbor pairs on the lattice, $H$ is an external field and $\Delta$ is the
strength of the disorder.  The random fields are independent random
variables chosen from a distribution with mean zero and variance one.  In
this study, the random fields are Gaussian and the lattice is simple cubic.

Currently, it is not known whether the phase transition for the 3D RFIM is
first-order or continuous.  Experiments on magnetic systems have been plagued
by problems of poor equilibration and have yielded confusing results, but it
appears that there is no latent heat at the transition and this has usually
been interpreted as evidence for a continuous transition.  Theoretical
analyses~\cite{Natt97} have also favored a continuous transition, although in
many cases this is an assumption rather than a conclusion, and some
 recent work~\cite{BrDe98} suggests a fluctuation driven first-order
transition.  The standard picture is that of a continuous transition
controlled by a zero temperature fixed point.  The scaling theory of this
transition~\cite{Villain85,BrMo,Fish86} has three independent critical
exponents and modified hyperscaling relations. Because of the fixed point
governing the transition has strong disorder, controlled renormalization group
calculations have not been possible. Migdal-Kadanoff renormalization group
calculations indicate a continuous transition~\cite{CaMa,FaBeMc} but also
mistakenly predict that the
$q$-state Potts transition is continuous for all $q$ when, for 3D it is known
to be first-order for $q \geq 3$.  Series analyses initially supported a
first-order transition~\cite{HoKhSe} but more recently point to a continuous
transition, at least for weak disorder~\cite{GoAdAhHaSc}.  Alternatively, it
may be that the transition is continuous for weak disorder and then becomes
first-order for strong disorder with a tricritical point separating the
critical line from the first-order line.

Recent Monte Carlo simulations~\cite{RiYo,Rieg95,NeBa} have also been
interpreted as showing a continuous transitions but with a jump in the
magnetization.  These simulations have been limited to system size
$16^{3}$.  The jump in the magnetization can be interpreted as a very small
value of the magnetization exponent but might also signal a first-order
transition.  Simulations of systems up to $64^3$~\cite{YoNa} were
interpreted as indicating a first-order transition but were clearly not
equilibrated in the transition region.  A number of numerical experiments
have also been carried out on the RFIM at zero temperature using polynomial
time ground state algorithms.  Originally these supported a continuous
transition with a small magnetization exponent~\cite{Ogielski} but more
recent studies on larger systems~\cite{Sour98} show a jump in the
magnetization and are thus suggestive of a first-order transition.

The numerical results presented in this paper differ from those of previous
studies in two significant respects.  First we use an efficient algorithm
that permits us to reach equilibrium for larger systems than those of past
studies.  Second, we fine-tune both the temperature and the external field
for each realization of disorder to be as close as possible to the
transition for that realization.

\begin{figure}[h]
\epsfxsize = 5in

\begin{center}
\leavevmode
\epsffile{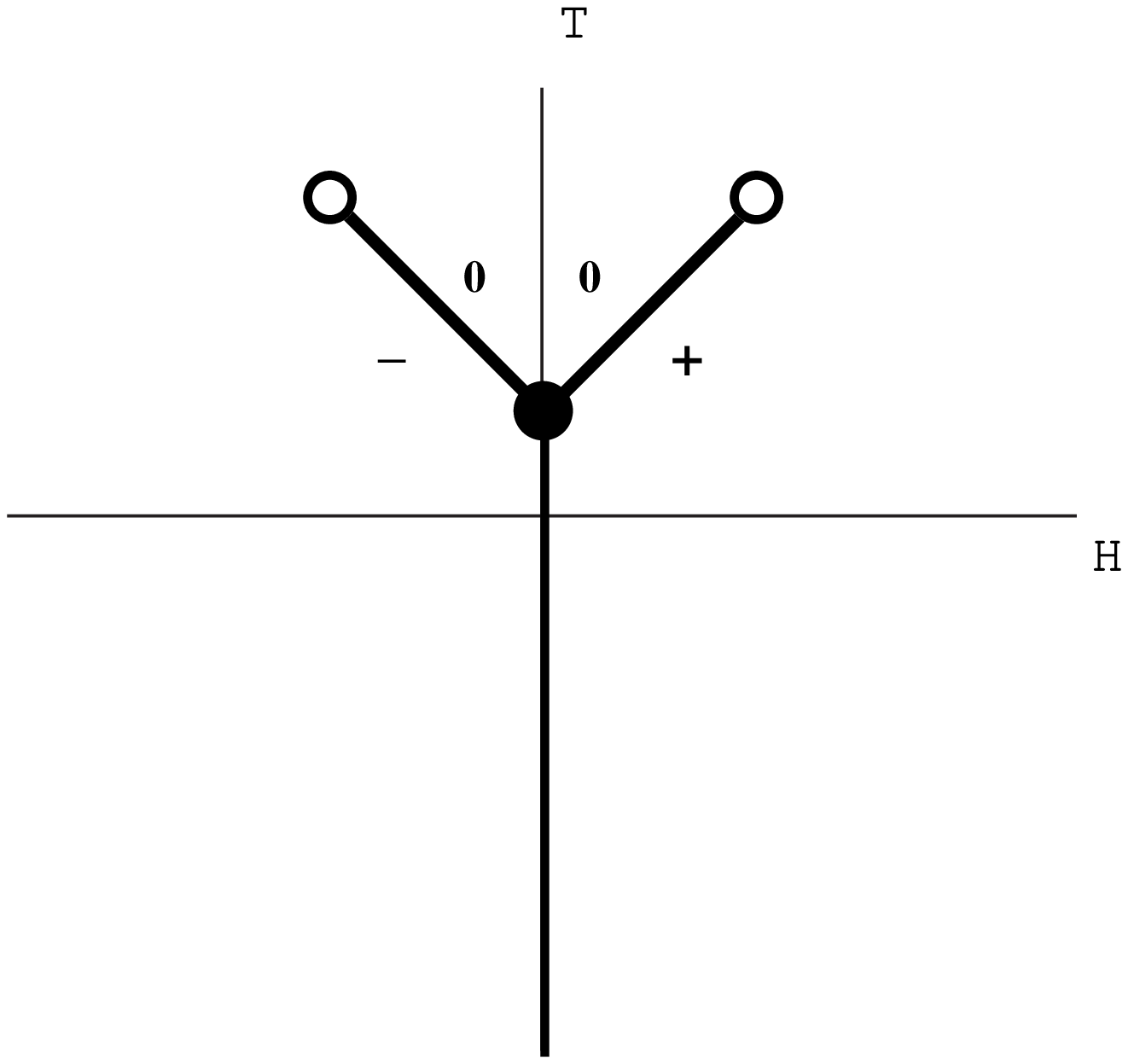}
\caption{Proposed phase diagram for the 3D RFIM in the $H$-$T$ plane  at a
  fixed strength of disorder.  The bold lines are first-order lines.  The
  black dot is the thermal first-order transition at ($T_c$,$H=0$) where
  the $+$,$-$ and $0$ phases coexist.  The open circles are critical
  endpoints of the two order-disorder first-order lines.}
\label{fig:phd}
\end{center}
\end{figure}

To understand the motivation for this fine-tuning, let us suppose for the
moment that the transition is first-order.  If this is the case, then for
periodic or helical boundary conditions we expect that there will be three
phases in coexistence at the transition point.  We call the coexisting
phases $+$, $-$ and $0$.  The ordered phases, $+$ and $-$, have long range
order and finite magnetization.  The disordered, $0$ phase has no long
range order, no magnetization and is characterized by spins that are
predominately aligned with their local fields.  The bond energy, defined by
the first term on the RHS of Eq.~(\ref{eq:define}), is greater in the $0$
phase than the ordered phases.  The expected phase diagram in the $T$--$H$
plane in the vicinity of the point of three phase coexistence is shown in
Fig.~\ref{fig:phd}.  For $T<T_c$ and $H=0$ there is phase coexistence of
the $+$ and $-$ phases that ends at the thermal first-order transition at
$T=T_c$ and $H=0$ (the black dot in the figure).  Since the $0$ and
$+$($-$) phases differ in both energy and magnetization, the disordered
phase can be maintained in coexistence with the $+$($-$) phase by
increasing the temperature and increasing (decreasing) the field.  The two
order--disorder lines corresponding to $0+$ and $0-$ coexistence form the
arms of the ``Y'' in the figure.

Since the disorder is independent and homogenous the free energy is self--averaging, 
hence, in the thermodynamic limit the transition occurs at a definite point ($T_c$, $H=0$)
for almost all realizations of the random fields.  However, for finite systems, the location
and qualitative features of the transition will depend on the realization of disorder.  In
particular, for a system of size $L$, with realization of random field, $\{h_i\}$ and
disorder strength $\Delta$ there may be a point where three phases coexist at
($T_c(\Delta;\{h_i\})$,
$H_c(\Delta;\{h_i\})$).  A first guess, based on the net field due to the
random fields, is that $T_c(\Delta;\{h_i\})$ and $H_c(\Delta;\{h_i\})$)
should be displaced from the average value $\tilde{T}_c(L)$ and $H=0$ by an
amount of order $\Delta L^{d/2}$.

How accurately must $H$ and $T$ be fine-tuned to see three phase
coexistence if it exists?  Suppose first that we want the $+$ and $-$
phases to coexist and that the magnetizations of these phases at the
transition differs by $2m$.
If the external field deviates from the correct value by $\delta H$, the
free energy difference between the $+$ and $-$ phase is $2 m \delta H L^d$.
For both phases to have a significant probability the free energy
difference must not greatly exceed $kT$.  Thus $H$ must be set to
$H_c(\Delta;\{h_i\})$ to within an accuracy of $kT/2 m L^{d}$ to have both
phases represented.  $H_c(\Delta;\{h_i\})$ will itself fluctuate from
sample to sample as $\Delta/L^{d/2}$.  Similarly, if the entropy difference
per spin between the the ordered and disordered phases is $s$, then $T$
must be fine-tuned to within $kT/sL^{d}$ to allow to these phases to
coexist.  Presumably sample to sample fluctuations in $T_c(\Delta;\{h_i\})$
also scale as $1/L^{d/2}$.  Thus if a single value of $H=0$ and
$T=\tilde{T}_c(L)$ is chosen for all realizations of the random field, one
will almost never see more than one phase at a time.


\section{Numerical methods}
\begin{figure}[h]
\epsfxsize = 5in
\begin{center}
\leavevmode
\epsffile{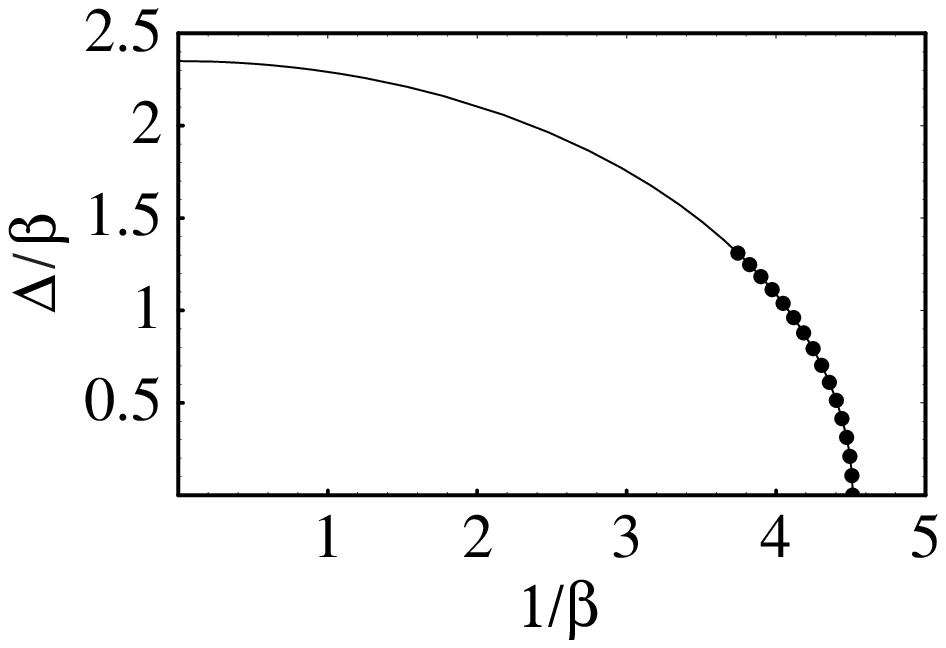}
\caption{Approximate phase diagram for the RFIM and replicas for the replica exchange
method.  The phase boundary in the $1/ \beta$--$\Delta/ \beta$ plane is taken to be an
elliptical curve starting at the pure Ising critical point ($1/\beta=4.512,\Delta/\beta=0$)
and ending at the zero temperature transition ($1/\beta=0,\Delta/\beta=2.35$).  The initial
conditions for the 16 replicas lie on this curve and are
evenly space in $\Delta$. During the feedback process, each replica is shifted by a small
amount in $\beta$ and $H$.}
\label{fig:phasediagram}
\end{center}
\end{figure}

We use an algorithm that combines the replica exchange method, first
introduced by Swendsen and Wang~\cite{SwWa86} and the two-replica cluster
method of Redner, Machta and Chayes~\cite{ReMaCh,ChMaRe98b}.  Our method is
also closely related to simulated and parallel tempering~\cite{MaPa}. The
idea of this approach is to simultaneously simulate $K$ replicas of the
system.  All replicas have the same normalized random field $\{h_{i}\}$ but
each replica has different values of the other parameters.  Replica $k$
($k=0, \ldots, K-1$) has inverse temperature $\beta_{k}$, strength of
randomness $\Delta_{k}$ and external field $H_k$.  The replicas form a
sequence so that neighboring replicas in the sequence are nearby in the
$(\beta,\Delta,H)$ parameter space.  Neighboring replicas exchange
magnetization with one another according to a procedure described below.
One end of the sequence of replicas is at a value of $\beta$, $\Delta$ and
$H$ that can be efficiently simulated using a known method while the other
end of the sequence is at a value of the parameters that we would like to
study.  In our case, the replicas lie along the RFIM phase transition line
starting from the pure Ising values $\beta_{0} \approx .22615$ and
$\Delta_{0}=H_{0} =0$ as shown in Fig.\ \ref{fig:phasediagram}.  The replicas are equally
spaced in $\Delta$.  The pure Ising replica is simulated using the Wolff single cluster
algorithm~\cite{Wolff,NeBa99}.  (We have also experimented with replicas
lying along a line of constant $\beta H$ starting at the RFIM phase
boundary, extending into the paramagnetic phase, and ending at a
temperature high enough that the model can be efficiently simulated using
the ordinary single-spin-flip Metropolis algorithm.  This approach,
however, is found to be less efficient than the one described above.)

Magnetization is exchanged between neighboring replicas using a
generalization of the two-replica cluster method.  Suppose we have two
replicas at $(\beta,\Delta, H)$ and $(\beta^{\prime},\Delta^{\prime},
H^{\prime})$ with spin configurations $\{S_i\}$ and $\{S_i^{\prime}\}$,
respectively.  A site $j$ is considered {\em active\/} 
for this pair of
replicas if $S_j \neq S_j^{\prime}$.  A bond $ij$ between neighboring sites
$i$ and $j$ is {\em satisfied\/} if $S_i = S_j$ and $S_i^{\prime}
=S_j^{\prime}$.  A cluster of active sites is formed starting from a
randomly chosen active site.  New active sites are added to the cluster by
{\em occupying\/} satisfied bonds on the perimeter of the cluster with
probability $p(\beta,\beta^{\prime})$ where
\begin{equation}
\label{eq:padd}
p(\beta,\beta^{\prime}) = 1 - e^{-2(\beta+\beta^{\prime})}.
\end{equation}
If a bond connecting a site to a cluster is occupied, the site is added to
the cluster and the set of bonds on the perimeter is updated.  In this way,
the cluster grows until no further sites are added.  The procedure is very
similar to the way clusters are grown in the Wolff single cluster
algorithm.

Once a cluster is identified it is {\em flipped\/} with a probability that
depends on the change in boundary and field energy so as to satisfy
detailed balance. 
 Flipping a cluster means changing the sign of all the
spins in the cluster or, equivalently, exchanging the values of the spins
in the cluster between the two replicas.  The probability to flip a cluster
${\cal C}$ with $\mid{\cal C}\mid$ sites depends on the quantity $\Sigma$
\begin{equation}
\label{eq:sigma}
\Sigma=-[(\Delta-\Delta^{\prime}) h_{\cal C} + (H-H^{\prime}) \mid {\cal C}
\mid + (\beta - \beta^{\prime})(N_{++}-N_{--})] S_{\cal C}
\end{equation}
where $N_{++}$ and $N_{--}$ are, respectively, the number of $++$ and $--$
sites that are nearest neighbors of the cluster, $h_{\cal C}$ is the net
random field acting on the cluster,
\begin{equation}
h_{\cal C}= \sum_{i \in {\cal C}} h_i,
\end{equation}
and $S_{\cal C}$ is the spin value of the cluster in the unprimed replica.
If $\Sigma \leq 0$ then the cluster is flipped, otherwise it is flipped
with probability $e^{-\Sigma}$.

It is straightforward but tedious to show that the choice of
$p(\beta,\beta^{\prime})$ (Eq.~\ref{eq:padd}) together with the flipping
probability defined by $\Sigma$ is precisely what is needed to ensure
detailed balance.  The motivation for the choice of
$p(\beta,\beta^{\prime})$ is most easily understand by considering the
limit where two replicas are at the same values of $\beta$, $\Delta$
and~$H$.  In that case $p(\beta,\beta) = 1- e^{-4\beta}$ and $\Sigma=0$.
Since $\Sigma=0$, clusters are always flipped just as is the case for the
Wolff single cluster algorithm.  Furthermore, we have shown in
Ref.~\onlinecite{ReMaCh,ChMaRe98b} that the active clusters percolate at
the RFIM phase transition.  If the transition is continuous clusters of all
sizes are flipped.  If the transition is first-order, there will be two
distinct kinds of clusters; some clusters will be extensive and change the
phase of the system while other clusters will have sizes less than or equal
to the correlation length.  In either case, the clusters identified by the
two-replica procedure correspond to the fluctuations that actually occur in
the system at the phase transition and permits large changes in the spin
configuration in a single Monte Carlo sweep.  When the two replicas do not
have equal values of the parameters, then the clusters do not flip freely
but if the replicas are close together in the parameter space, the
acceptance fraction for flipping clusters will remain high.

Our method and the original replica exchange method~\cite{SwWa86} on which it
is based is similar to parallel tempering~\cite{MaPa}.  In all these methods,
groups of spins are exchanged between neighboring replicas along a sequence.  In
parallel tempering the whole spin configuration is exchanged and the Boltzmann
factor controlling the acceptance of the move depends on the energy difference
between the replicas.  In our algorithm, only some of the spins are exchanged
and for a given distance in the parameter space between the replicas the
acceptance fraction is larger than for parallel tempering.  The consequence is
that fewer, less closely spaced replicas are needed 
for the replica exchange method.


In order to find the phase transition temperature and external field
($\beta_c(\Delta;\{h_i\})$, $H_c(\Delta;\{h_i\})$) for a given realization
and strength of disorder we use a feedback mechanism.  Starting from an
initial value of $\beta$ and $H$ we monitor the magnetization of the system
after each Monte Carlo sweep.  If the absolute value of the magnetization
is less than a lower cut-off $M_{lc}$ the system is interpreted to be in a
high temperature phase and the inverse temperature is increased by a small
amount~$\epsilon_\beta$.  If the absolute value of the magnetization is
greater than an upper cut-off $M_{uc}$ the system is interpreted to be in
one of the ordered phases and $\beta$ is decreased by $\epsilon_\beta$.  In
this case, the external field is also adjusted by an amount $\epsilon_H$ if
the magnetization is positive and by $-\epsilon_H$ if the magnetization is
negative.  The
 average value of $\beta$ and $H$ is computed for the period
when the feedback procedure is on and is taken to be
($\beta_c(\Delta;\{h_i\})$, $H_c(\Delta;\{h_i\})$).  The feedback procedure
is then turned off and these values are used for a long equilibrium
simulation.

Most of our simulations were for $24^3$ systems.  Except as otherwise
noted, the simulations used $K=16$ replicas equally spaced in $\Delta$ with
the most disordered replica having $\Delta=.35$.  Initially the replicas lie on an
elliptical curve in the temperature-disorder plane that starts at the pure Ising transition
($\beta=0.22165$) and ends at the zero temperature transition ($\Delta/\beta
=2.35$) as shown in Fig.\
\ref{fig:phasediagram}.  It was shown in Ref.\
\onlinecite{NeBa} that this curve is a good estimate for the phase boundary.  
The feedback parameters were
$M_{lc}=2000$,
$M_{uc}=4000$,
$\epsilon_\beta=10^{-4}$ and
$\epsilon_H=10^{-5}$.  The feedback mechanism was run for $2 \times 10^5$
Monte Carlo sweeps and then, with the temperature and external field
determined by the feedback procedure, data were collected for $8 \times
10^5$ Monte Carlo sweeps.  Each Monte Carlo sweep is defined as 500 cluster
steps; since the average cluster size was found to be roughly 1000, a
single Monte Carlo sweep corresponds to attempting to flip every spin in
every replica about once.  For each cluster step a random integer $k$ was
chosen between 0 and $K$.  If $k=0$, a Wolff move was performed on the pure
replica, if $1<k<K$, a two replica cluster move was carried out between
replica $k$ and replica~$k-1$.  If $k=K$, a two replica cluster move was
carried out between replicas $K-1$ and~$K-2$.  The procedure described
above, totaling $10^6$ sweeps, was performed for 22 realizations of
disorder, a number chosen in advance of the experiment, labeled 10 through 31 by
the seed of the random number generator that produced $\{h_i\}$.  Each
realization required about six days on a 450MHz Pentium III machine.  A number
of additional simulations were carried out for specific realizations of
disorder (not all in the range 10 through 31) using parameters that differed in
some way from the above.  In particular, we did a careful study of realization
1 that is described below.

\section{Results}

\subsection{Energy and magnetization histograms}

\begin{figure}[h]
\epsfxsize = 5in
\begin{center}
\leavevmode
\epsffile{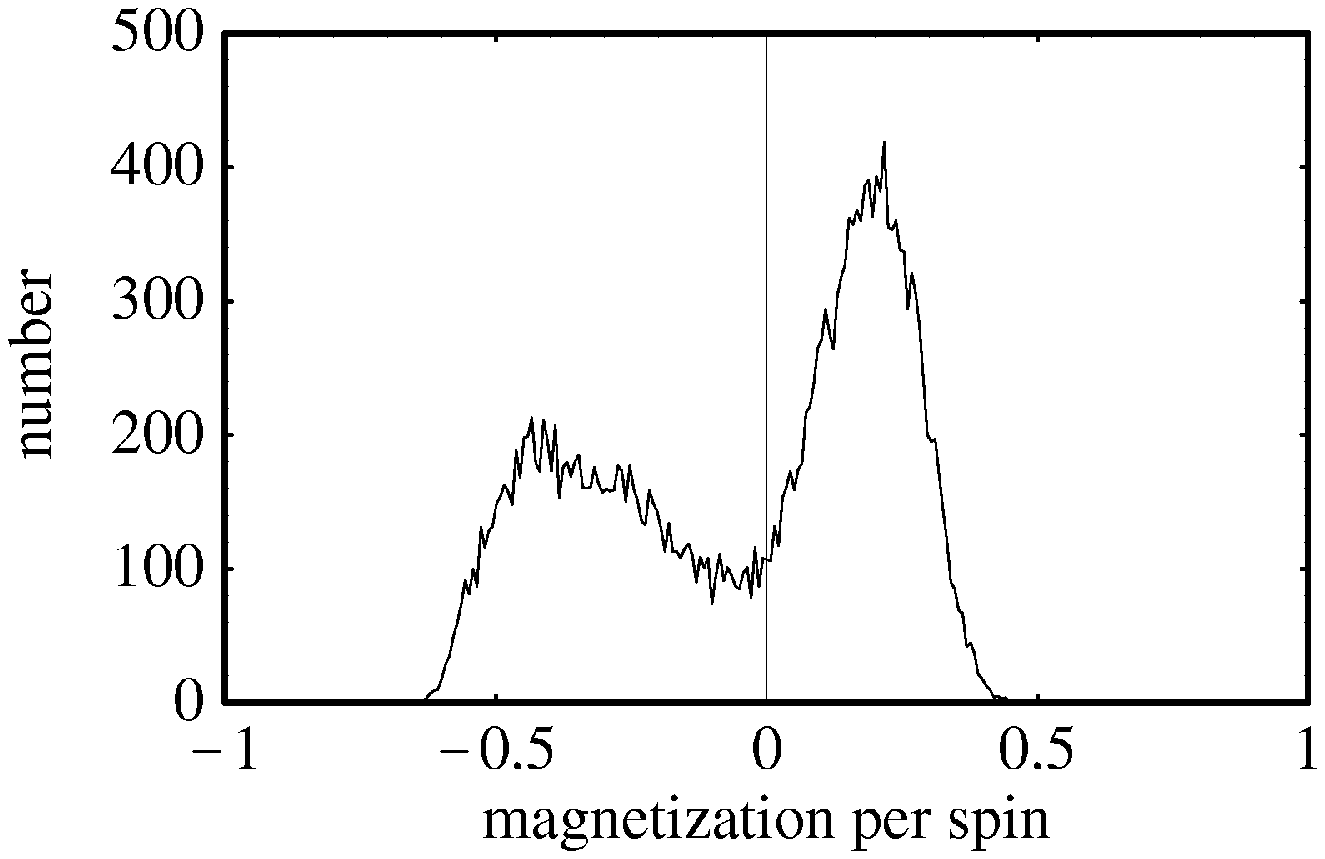}
\caption{Magnetization histogram for realization 20 at $\Delta=0.35$.}
\label{fig:m20}
\end{center}
\end{figure}

\begin{figure}[h]
\epsfxsize = 5in
\begin{center}
\leavevmode
\epsffile{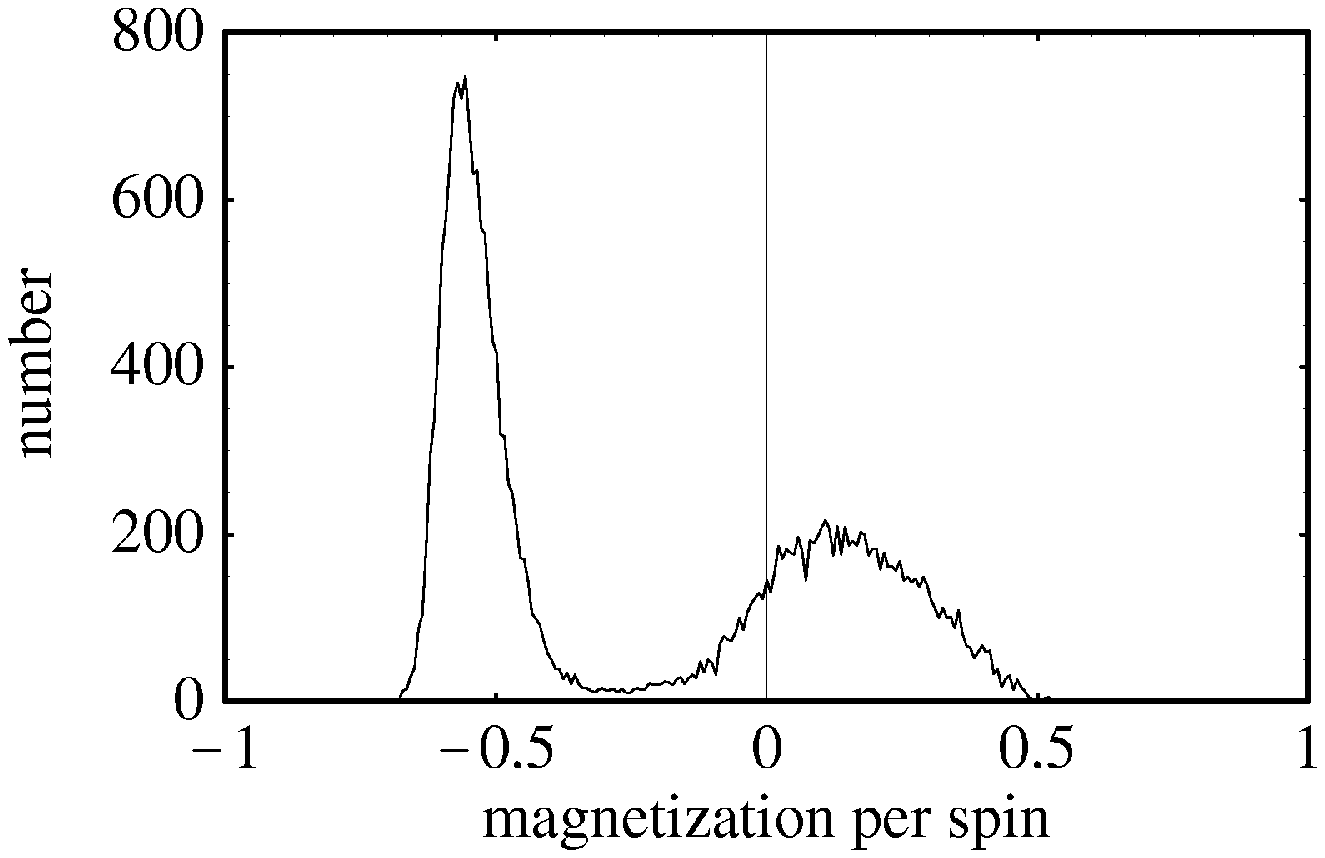}
\caption{Magnetization histogram for realization 21 at $\Delta=0.35$.}
\label{fig:m21}
\end{center}
\end{figure}

\begin{figure}[h]
\epsfxsize = 5in
\begin{center}
\leavevmode
\epsffile{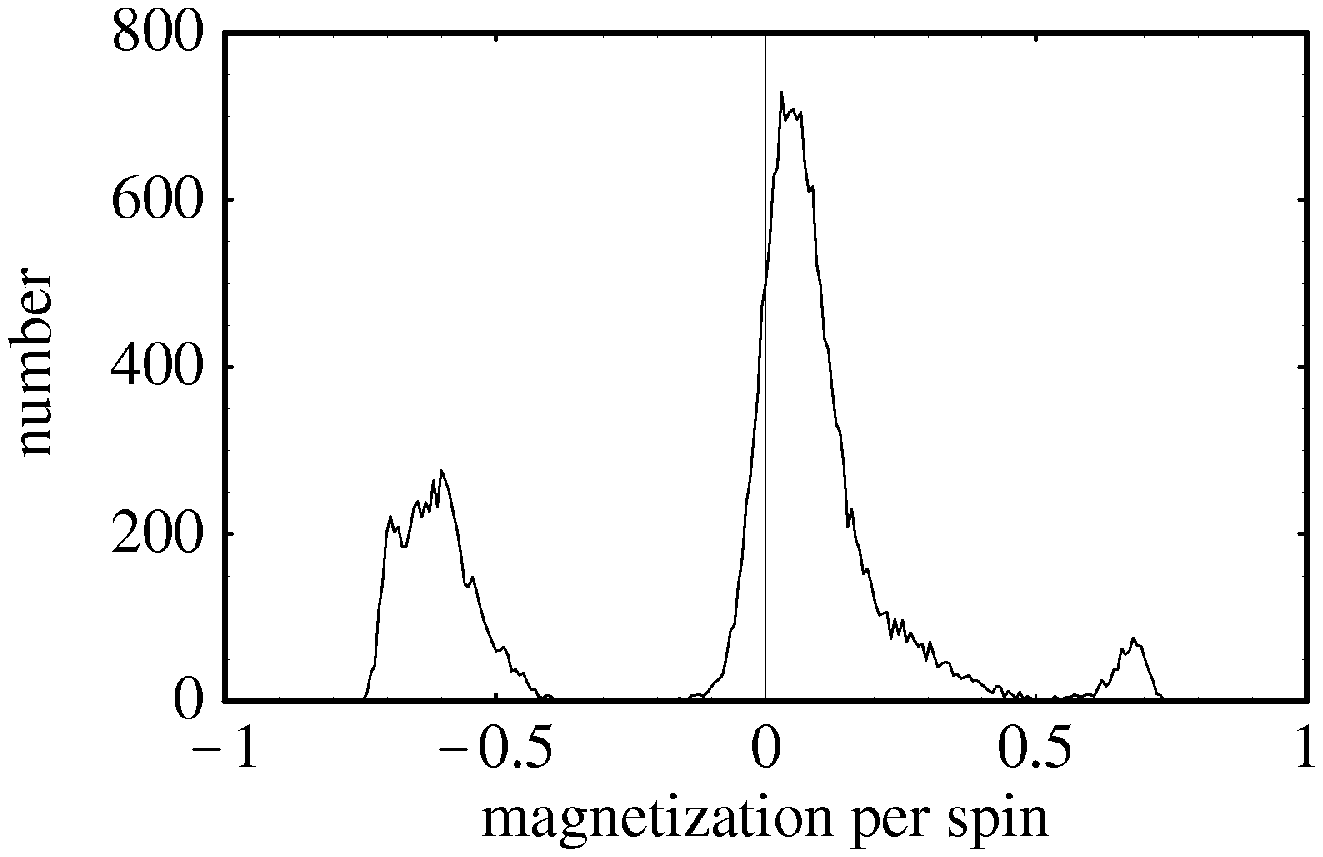}
\caption{Magnetization histogram for realization 25 and $\Delta=0.35$.}
\label{fig:m25}
\end{center}
\end{figure}

\begin{figure}[h]
\epsfxsize = 5in
\begin{center}
\leavevmode
\epsffile{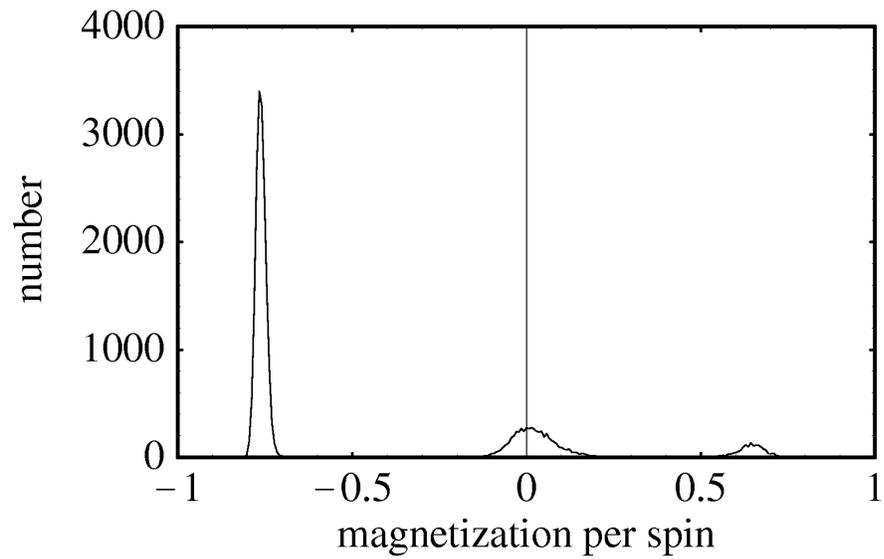}
\caption{Magnetization histogram for realization 31 at $\Delta=0.3267$.}
\label{fig:m31}
\end{center}
\end{figure}

\begin{figure}[h]
\epsfxsize = 5in
\begin{center}
\leavevmode
\epsffile{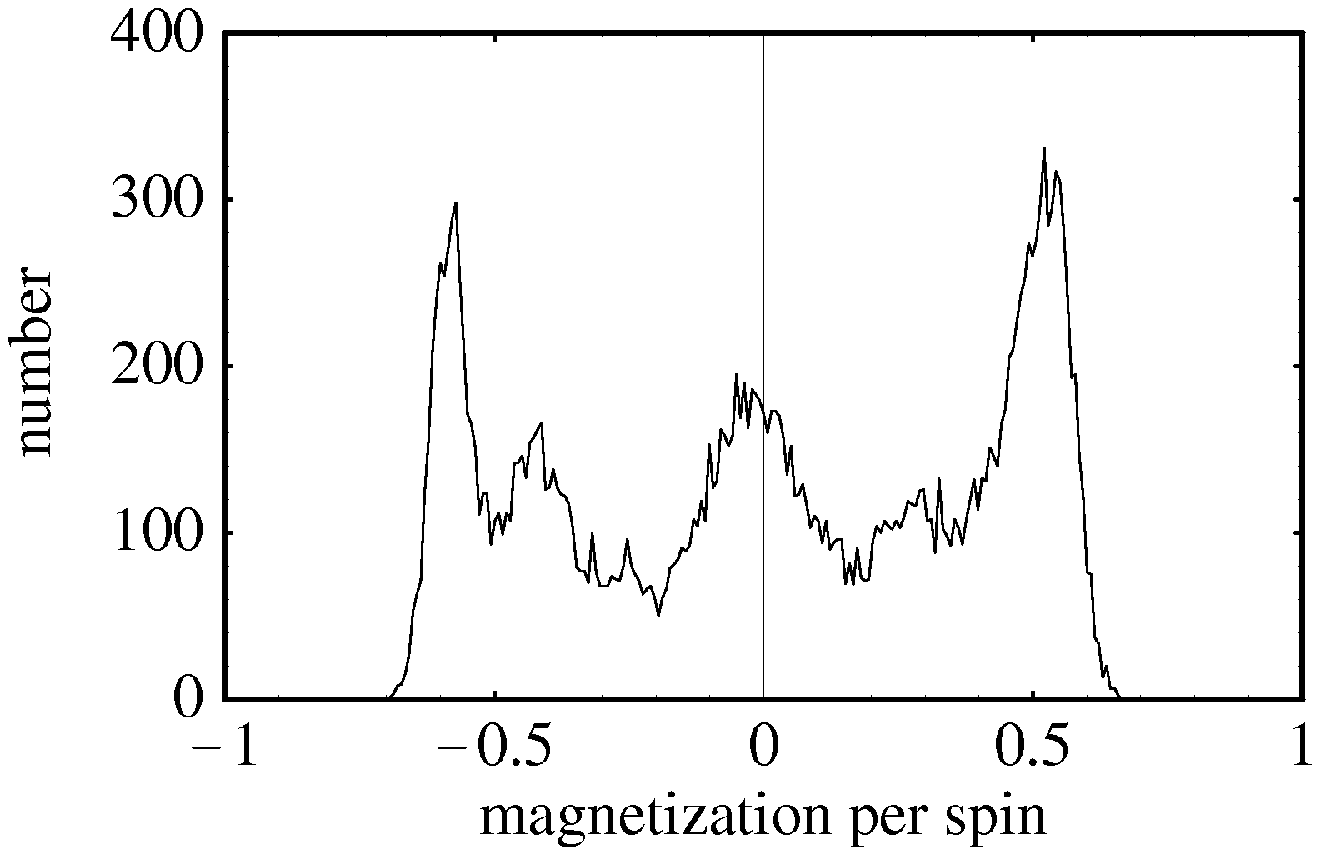}
\caption{Magnetization histogram for realization 14 and $\Delta=0.35$.}
\label{fig:m14}
\end{center}
\end{figure}

\begin{figure}[h]
\epsfxsize = 5in
\begin{center}
\leavevmode
\epsffile{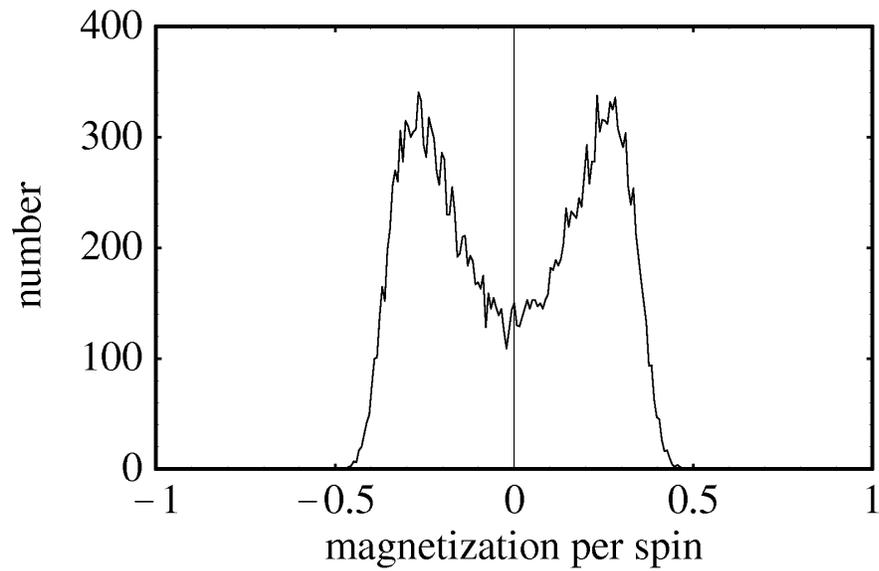}
\caption{Magnetization histogram for the pure Ising model at the infinite system
size critical
  temperature $\beta_c=0.22165$.}
\label{fig:mpure}
\end{center}
\end{figure}

The main lesson of our work is that each realization of disorder has its
own character.  This is best seen by examining the probability
distributions of magnetization and energy for several realizations of
disorder.  Figures~\ref{fig:m20}, \ref{fig:m21}, \ref{fig:m25},
\ref{fig:m31} and~\ref{fig:m14} show the magnetization histogram for
realizations 20, 21, 25, 31 and~14, respectively.  The magnetization
histogram for realization~20, characterized by two broad maxima roughly
symmetrical about the origin, is typical of the majority of realizations.
Apart from the asymmetry in the two peaks, realization~20 and those like it
are similar to the pure Ising model whose magnetization histogram is shown
in Fig.~\ref{fig:mpure} at the infinite system size critical temperature,
$\beta_c(0)=.22165$.  The magnetization histogram for realization~21,
Fig.~\ref{fig:m21} also displays two peaks but now one peak is much sharper
than the other and quite asymmetric about the origin.  The magnetization
histogram for realizations~25, Fig.~\ref{fig:m25} and~31,
Fig.~\ref{fig:m31} are qualitatively different, displaying three distinct
and well separated peaks.  It should be noted that the feedback procedure
was initially unsuccessful for these realizations and had to be run again
with $M_{lc}=M_{uc}=5000$ to find all three peaks.  Failure of the feedback
scheme was evidenced by the presence of only one narrow peak in the
magnetization histogram.  In the case of realization~31, the feedback procedure
with the revised parameters failed on the most disordered replica so
Fig.~\ref{fig:m31} shows the second to last replica in the sequence at
$\Delta=.3267$.  For both realization 25 and
31, the original run showed three peaked structures at weaker disorder giving a
strong hint that the same qualitative feature would be found at stronger
disorder. Finally, realization~14, Fig.~\ref{fig:m14} was unique, displaying a
five peaked structure.  Out of a total of 22 magnetization histograms, 19 show
two peaks, 2 three peaks and 1 five peaks.  It is possible but unlikely that
some of the two peak realizations would display more peaks with different
fine-tuning.  For example, we used other alternate values
 of the
fine-tuning parameters to search for an additional positive peak for
realization 21 but none was found.

\begin{figure}[h]
\epsfxsize = 5in
\begin{center}
\leavevmode
\epsffile{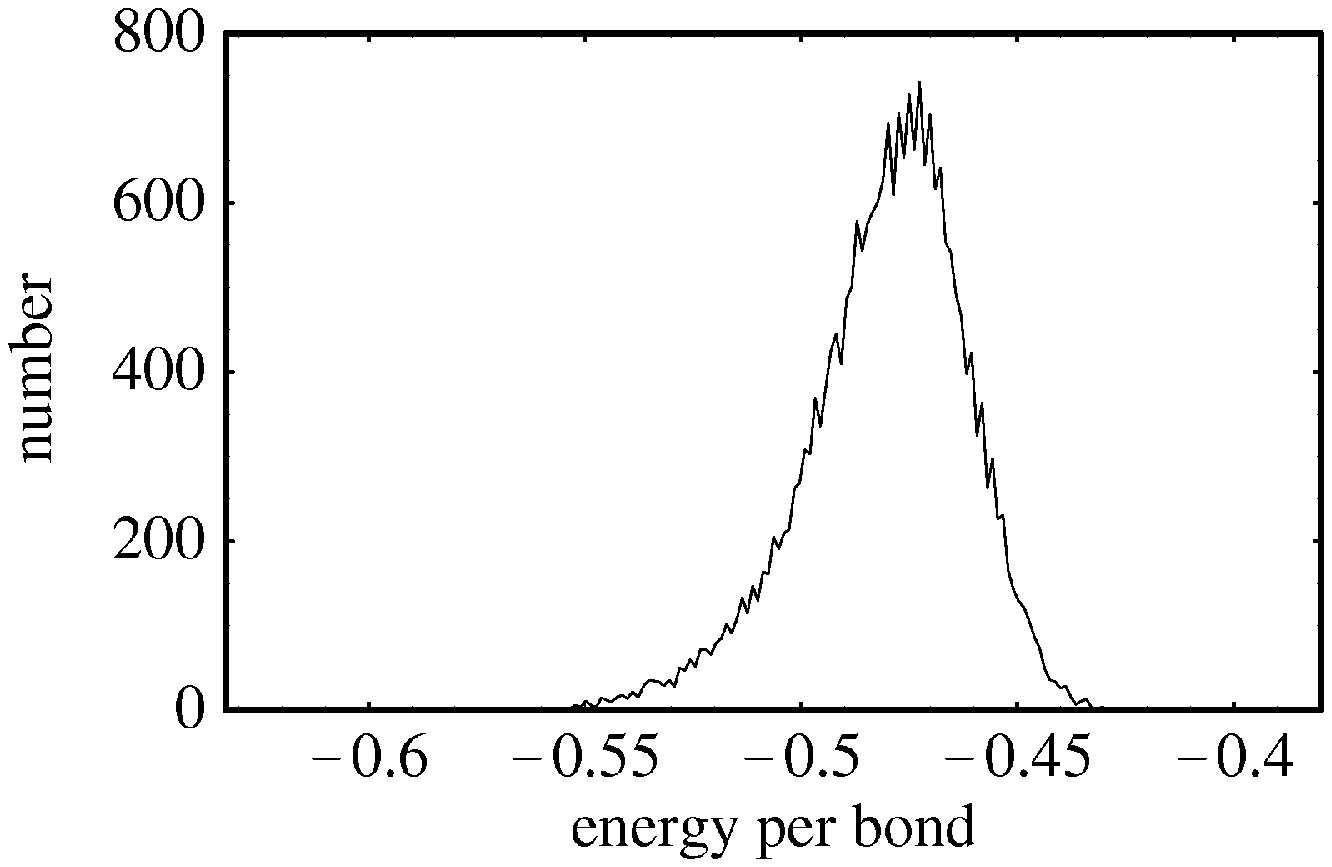}
\caption{Bond energy histogram for realization 20 at $\Delta=0.35$.}
\label{fig:e20}
\end{center}
\end{figure}

\begin{figure}[h]
\epsfxsize = 5in
\begin{center}
\leavevmode
\epsffile{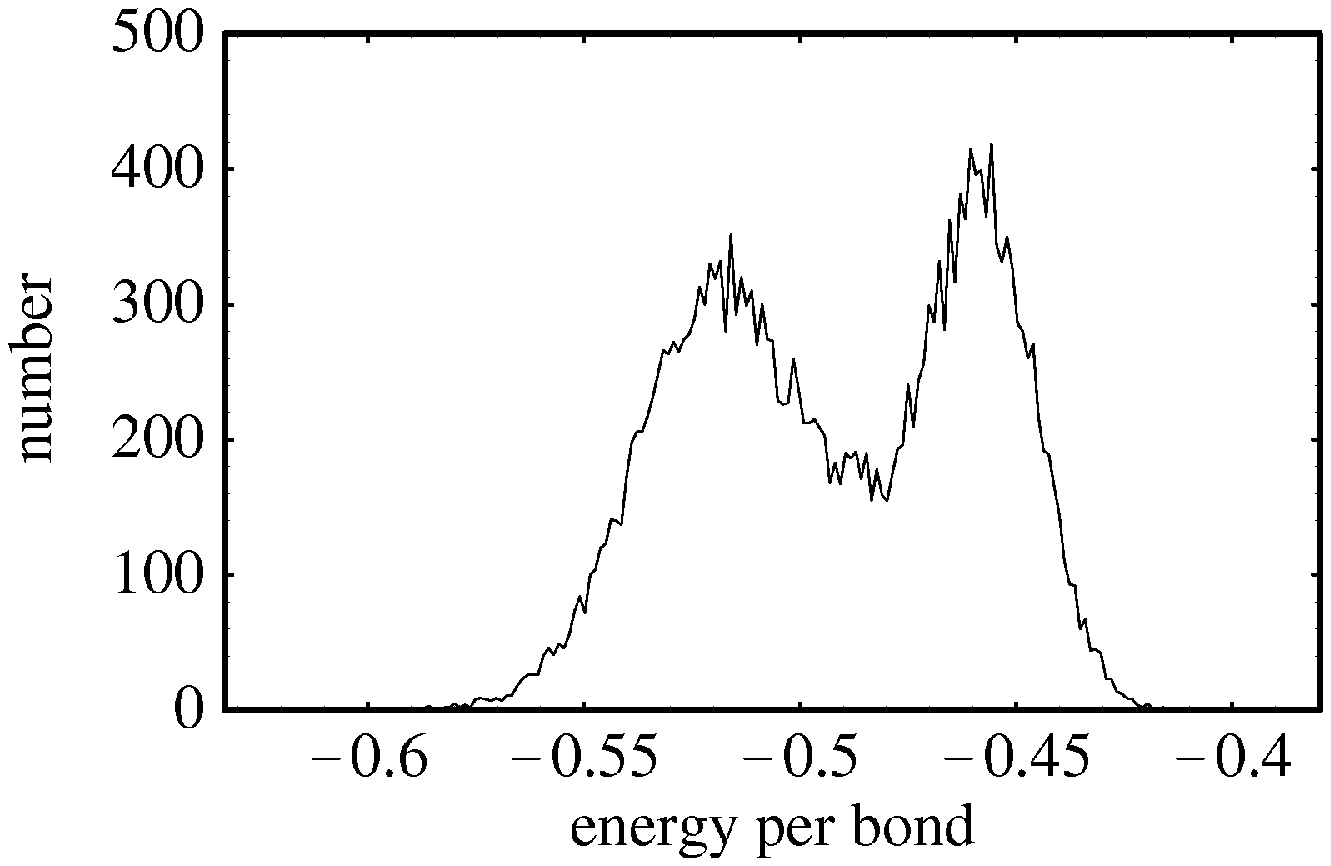}
\caption{Bond energy histogram for realization 21 at $\Delta=0.35$.}
\label{fig:e21}
\end{center}
\end{figure}

\begin{figure}[h]
\epsfxsize = 5in
\begin{center}
\leavevmode
\epsffile{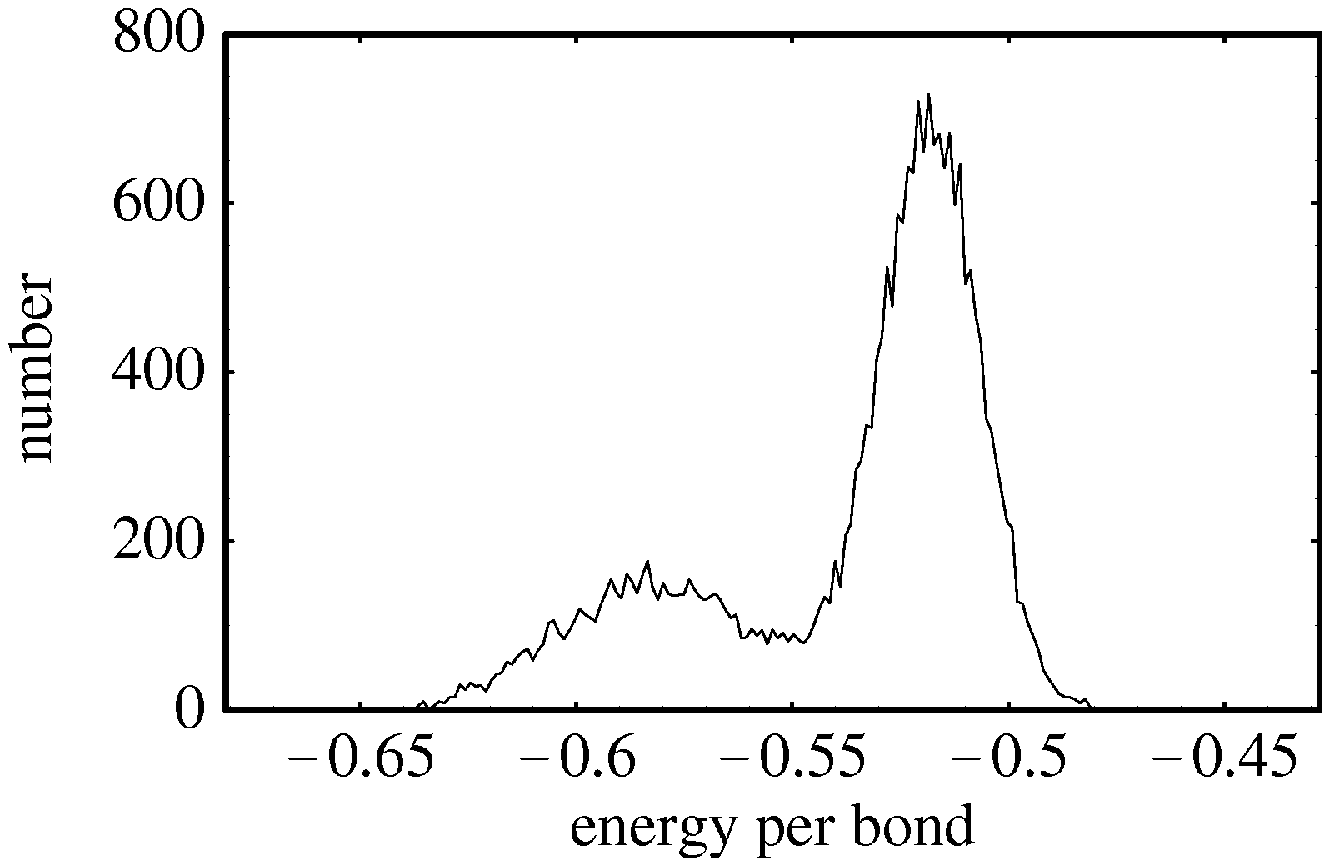}
\caption{Bond energy histogram for realization 25 at $\Delta=0.35$.}
\label{fig:e25}
\end{center}
\end{figure}

\begin{figure}[h]
\epsfxsize = 5in
\begin{center}
\leavevmode
\epsffile{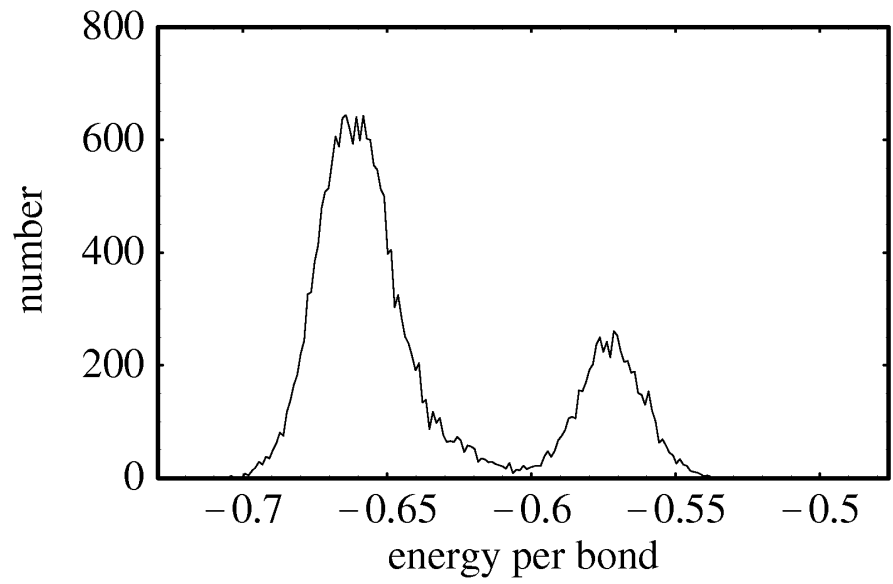}
\caption{Bond energy histogram for realization 31 at $\Delta=0.3267$.}
\label{fig:e31}
\end{center}
\end{figure}

\begin{figure}[h]
\epsfxsize = 5in
\begin{center}
\leavevmode
\epsffile{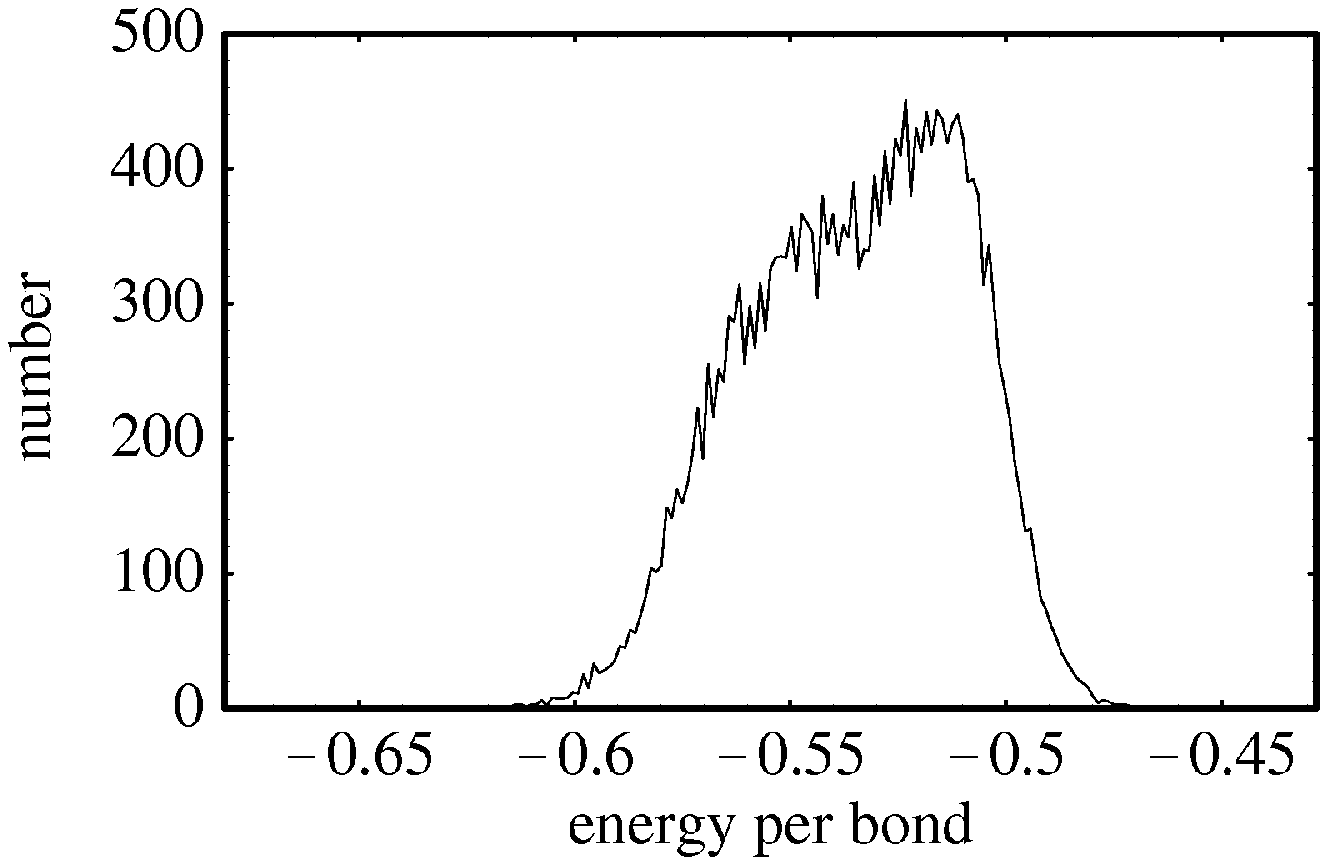}
\caption{Bond energy histogram for realization 14 at $\Delta=0.35$.}
\label{fig:e14}
\end{center}
\end{figure}

\begin{figure}[h]
\epsfxsize = 5in
\begin{center}
\leavevmode
\epsffile{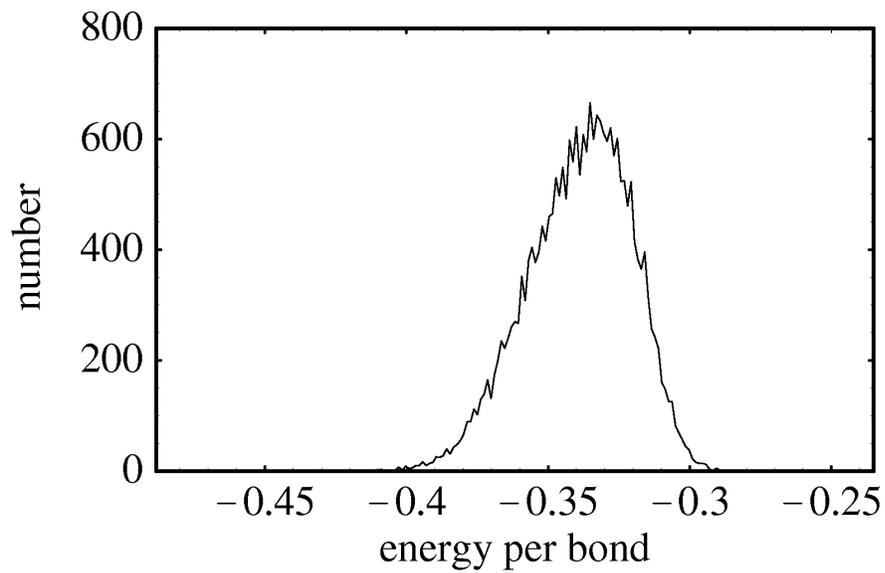}
\caption{Bond energy histogram for the pure Ising model at the infinite
  system size critical temperature $\beta_c=0.22165$.}
\label{fig:epure}
\end{center}
\end{figure}

The bond energy distribution displayed two possible qualitative behaviors.
Figures~\ref{fig:e20}, \ref{fig:e21}, \ref{fig:e25}, \ref{fig:e31}
and~\ref{fig:e14} show the bond energy histogram for realizations~20, 21,
25, 31 and~14, respectively.  For comparison, Fig.~\ref{fig:epure} shows
the energy distribution for the pure Ising model at the infinite system
size critical temperature.  For realization~20 and those like it, the bond
energy distribution has a single peak and is qualitatively like that of the
pure Ising model at criticality, except that the peak is shifted to
significantly lower energies.  For realization~21, which has two well
separated magnetization peaks, there are also two well separated peaks in
the bond energy distribution.  Similarly, for realizations~25 and~31 there
are two peaks in the bond energy distribution.  In a first-order
interpretation
 of these realizations the $+$ and $-$ phases have the lower
energy and the $0$ phase the higher.  On the other hand, for
realization~14, which has five magnetization peaks, there is a single broad
maximum in the bond energy histogram with a significant, low energy
shoulder.  Altogether, 5 out of 22 realizations show two well defined peaks
in the bond energy histogram while the remaining realizations show a single
peak though sometimes with a significant low energy shoulder.

\begin{figure}[h]
\epsfxsize = 5in
\begin{center}
\leavevmode
\epsffile{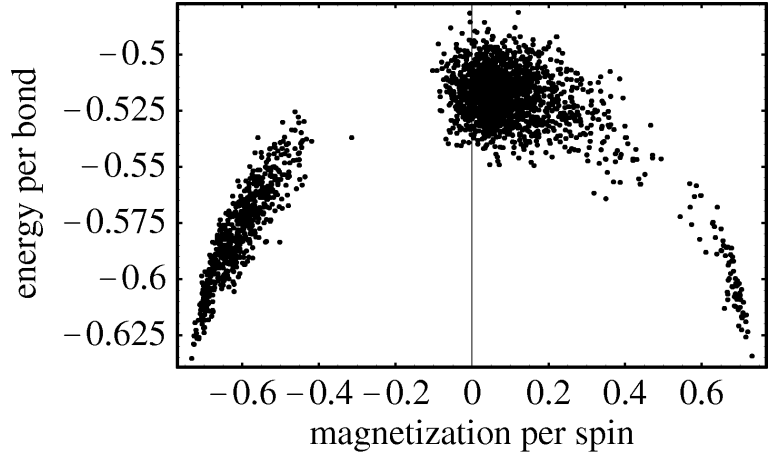}
\caption{The joint magnetization/bond-energy distribution for
  realization~25 at $\Delta=0.35$.}
\label{fig:joint25}
\end{center}
\end{figure}

\begin{figure}[h]
\epsfxsize = 5in
\begin{center}
\leavevmode
\epsffile{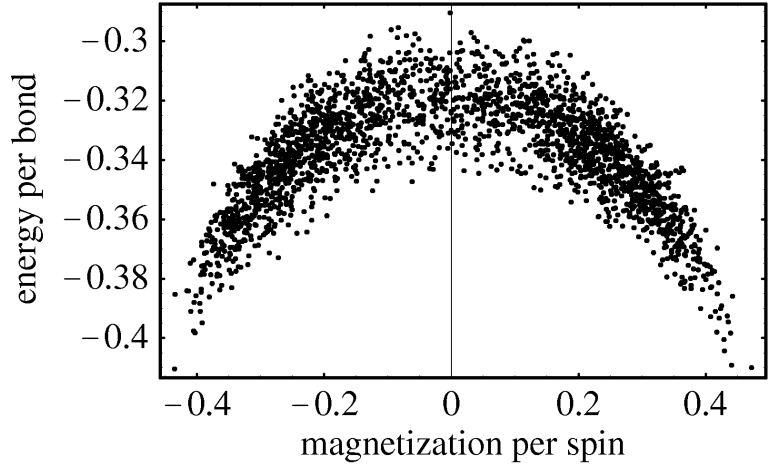}
\caption{The joint magnetization/bond-energy distribution for for the
  pure Ising model at the infinite system size critical temperature
  $\beta_c=0.22165$..}
\label{fig:jointpure}
\end{center}
\end{figure}

The joint magnetization/bond-energy distribution for realization~25 is
shown in Fig.~\ref{fig:joint25}.  For comparison, the joint distribution
for the pure Ising model at the infinite system critical temperature is
shown in Fig.~\ref{fig:jointpure}.  The three lobes in the distribution
correspond to the three peaks in the magnetization histogram,
Fig.~\ref{fig:m25}.  It is tempting to interpret realizations 25 and 31 in
the language of first-order transitions and declare the low energy side
lobes as ordered phases and the central lobe as a disordered phase.

\begin{figure}[h]
\epsfxsize = 5in
\begin{center}
\leavevmode
\epsffile{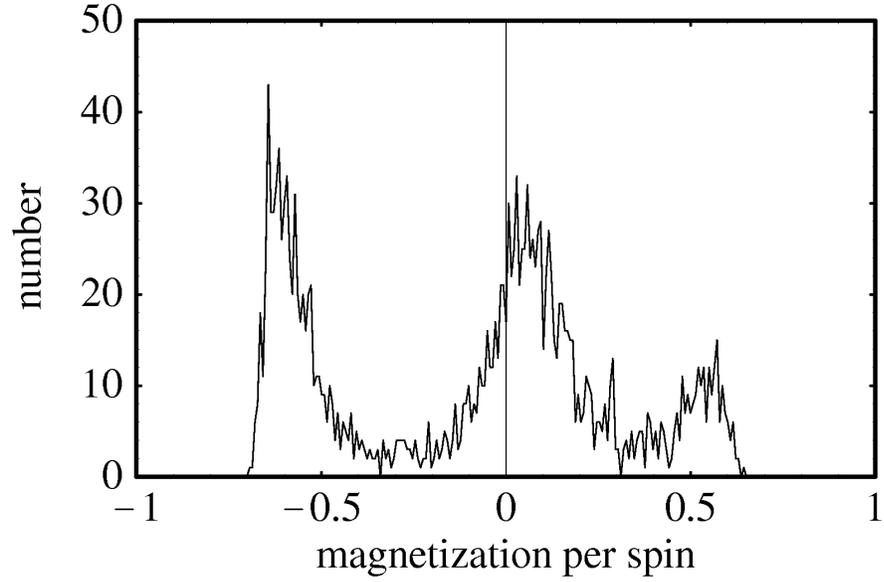}
\caption{Magnetization histogram for realization~1 at $\Delta=0.35$ and
  the phase transition point determined from fine-tuning,
  $\beta_c=0.268385$ and $H_c=0.00127049$.}
\label{fig:m1}
\end{center}
\end{figure}

\begin{figure}[h]
\epsfxsize = 5in
\begin{center}
\leavevmode
\epsffile{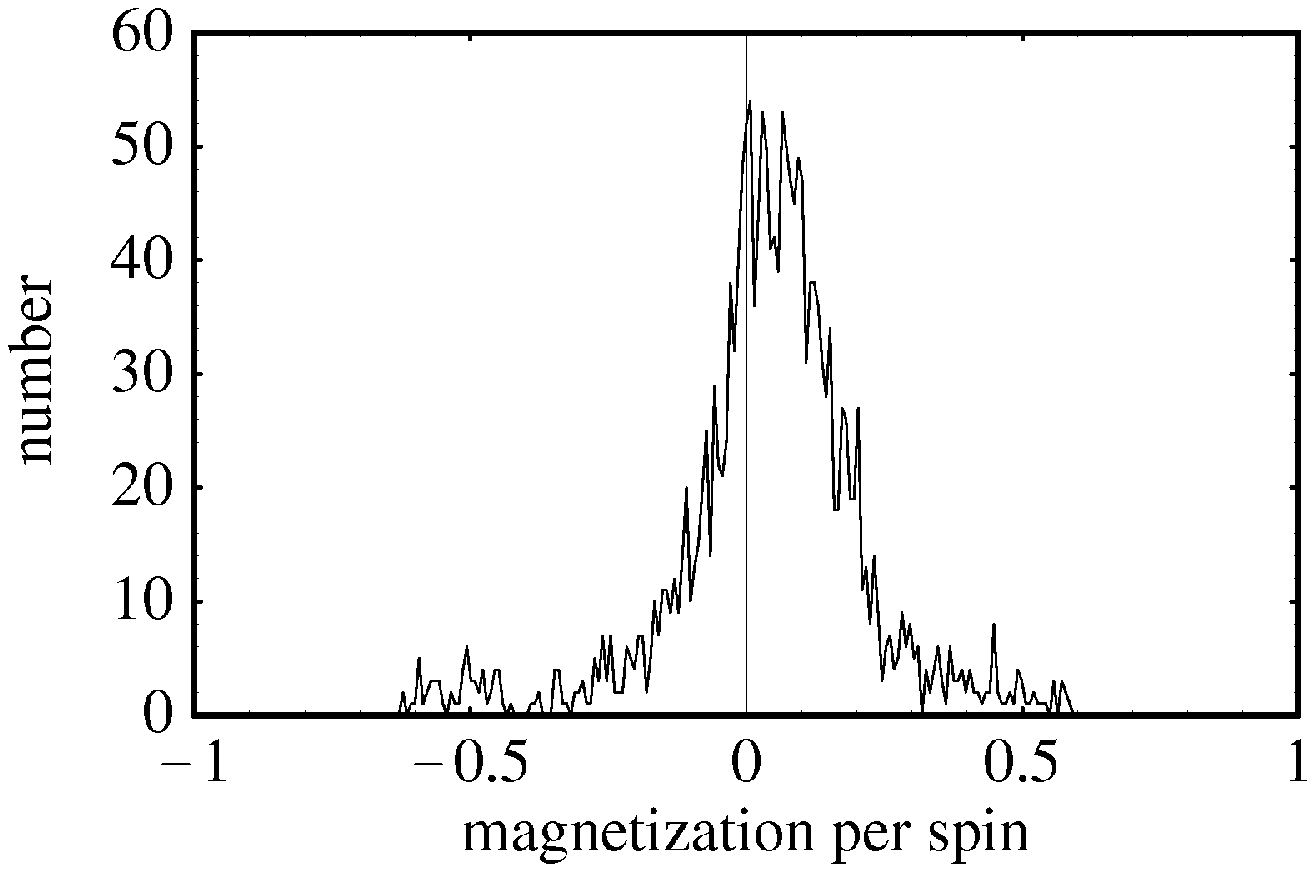}
\caption{Magnetization histogram for realization~1 at $\Delta=0.35$ and 
  $\beta=0.267041$ ($5\%$ warmer than $\beta_c$ determined for realization
  1) and $H_c=0.00127049$.}
\label{fig:m1h}
\end{center}
\end{figure}

\begin{figure}[h]
\epsfxsize = 5in
\begin{center}
\leavevmode
\epsffile{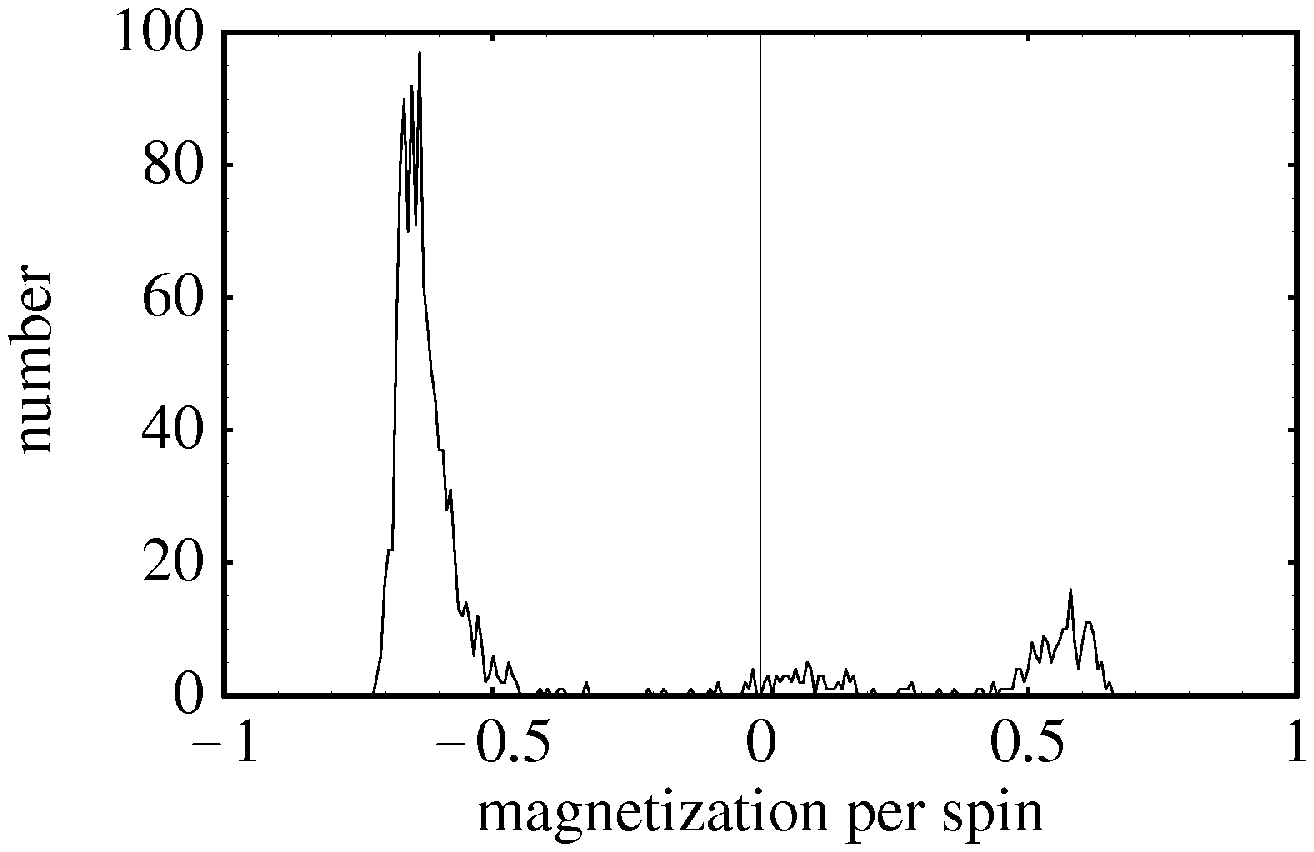}
\caption{Magnetization histogram for realization 1 at $\Delta=0.35$ and 
  $\beta=0.269727$ ($5\%$ colder than $\beta_c$ determined for realization
  1) and $H_c=0.00127049$.}
\label{fig:m1c}
\end{center}
\end{figure}

\begin{figure}[h]
\epsfxsize = 5in
\begin{center}
\leavevmode
\epsffile{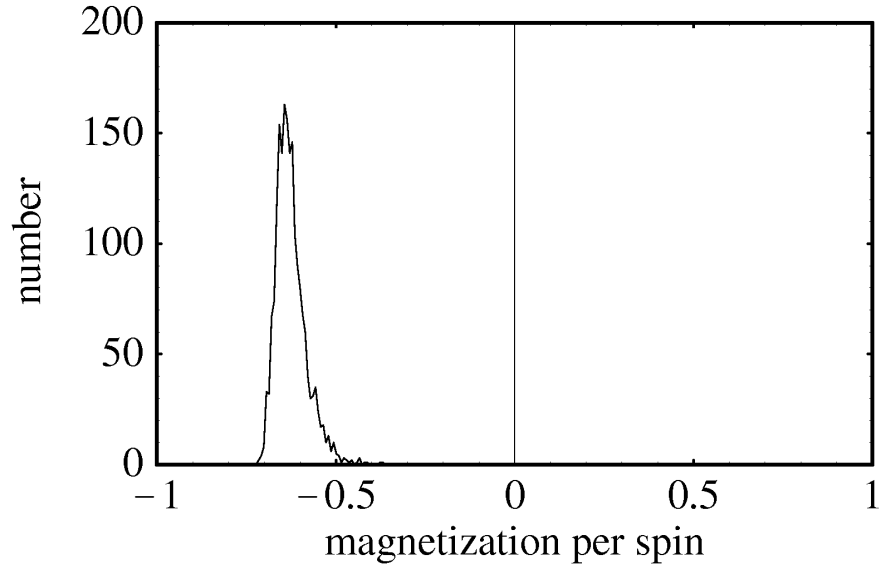}
\caption{Magnetization histogram for realization 1 at $\Delta=0.35$, 
  $\beta_c=0.267041$ and $H=0$.}
\label{fig:m1H0}
\end{center}
\end{figure}

The importance of fine-tuning the magnetization and temperature for each
realization of disorder is illustrated in Figs.~\ref{fig:m1},
\ref{fig:m1h}, \ref{fig:m1c} and~\ref{fig:m1H0}.  The magnetization
histogram for realization~1 at the fine-tuned parameter values,
$\beta_c=0.268385$ and $H_c=0.00127049$ is shown in Fig.~\ref{fig:m1}.
Note, that this value of $H_c$ is not the same as the value that would be
obtained by summing the local fields.  For realization one, that value is
almost $50\%$ larger, $-\Delta \sum_i h_i/N=0.00178774$.
Figure~\ref{fig:m1h} shows realization~1 with same external field but with
the temperature raised by $5\%$ leaving only the disordered phases.
Figure~\ref{fig:m1c} shows realization~1 with the same external field but
the temperature decreased by~$5\%$.  Although, the external field is too
negative here, it is nonetheless clear that the two ordered phases are now
dominant over the disordered phase, which has nearly vanished.  Finally, if
the external field is set to 0 and the feedback temperature is used, only
the negative peak survives, as shown in Fig.~\ref{fig:m1H0}.

\begin{figure}[h]
\epsfxsize = 5in
\begin{center}
\leavevmode
\epsffile{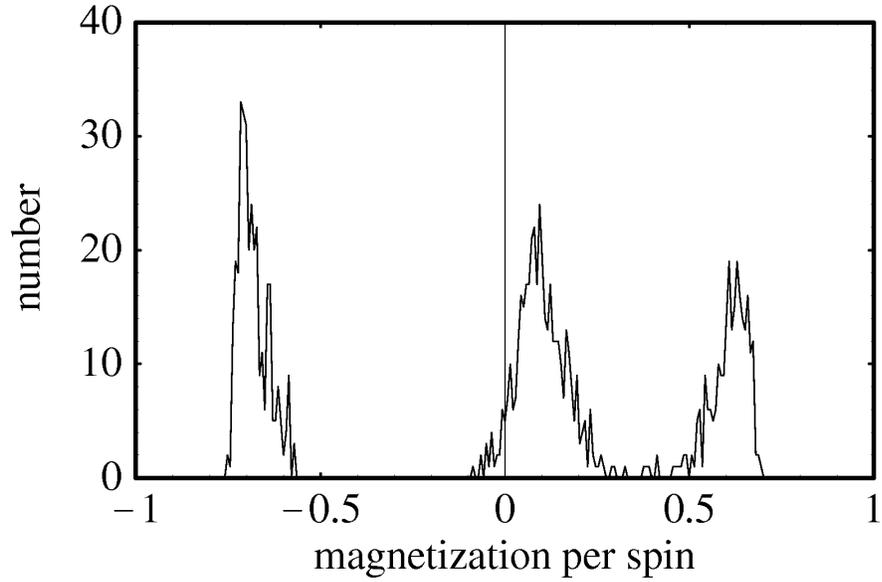}
\caption{Magnetization histogram for realization 1 at $\Delta=0.433$ at
  temperature and external field determined from fine-tuning,
  $\beta_c=0.289176$ and $H_c=0.00158005$.}
\label{fig:m1s}
\end{center}
\end{figure}

\begin{figure}[h]
\epsfxsize = 5in
\begin{center}
\leavevmode
\epsffile{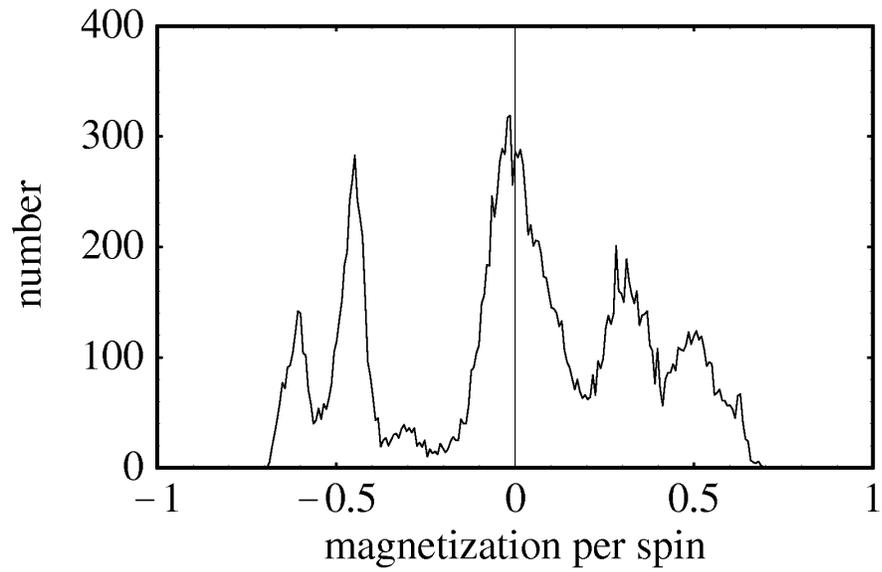}
\caption{Magnetization histogram for realization 14 at $\Delta=0.5$ at
  temperature and external field determined from fine-tuning.}
\label{fig:m14s}
\end{center}
\end{figure}

We considered several realizations at stronger disorder.
Figure~\ref{fig:m1s} shows the magnetization histogram for realization~1 at
disorder $\Delta=0.433$ at the value of external field and temperature
determined from the fine-tuning procedure.  Comparing Fig.~\ref{fig:m1s} to
Fig.~\ref{fig:m1} we see that the three-peaked structure is conserved and
the peaks becomes sharper as the strength of the disorder is increased.  In
this experiment, we used 16 replicas with maximum disorder strength,
$\Delta=0.5$ however, the fine-tuning procedure failed for the two most
disordered replicas and only yielded the minus phase.  In general, the
feedback procedure was less stable for stronger disorder.
Figure~\ref{fig:m14s} shows realization 14 at disorder $\Delta=0.5$ at the
values of the temperature and 
external field determined from the
fine-tuning procedure.  A comparison to Fig.~\ref{fig:m14} shows that
increasing the disorder has sharpened the five peaked structure of the
magnetization histogram.  It would be interesting to carry out a systematic
study of stronger disorder to see if there is a trend toward more
``first-order'' like realization however this will require developing a
better fine-tuning mechanism.

Overall, out of 21 realizations (seeds 10-30) at disorder strength $\Delta=.35$
we found that the average values of the critical parameters is $\langle \beta_c
\rangle = 0.2663$ and $\langle H_c \rangle=0.0006 $ (compared with initial values before the
feedback procedure of $\beta=0.2670$ and $H=0$) with a standard deviations of
$\sigma_{\beta} = 0.0037$ and $\sigma_{H} = 0.0031$. The standard deviation of $H_c$ is
consistent with the anticipated value $\sigma_H \sim
\Delta L^{-d/2} = 0.0030$.

\subsection{Dynamics of the algorithm}

\begin{figure}[h]
\epsfxsize = 5in
\begin{center}
\leavevmode
\epsffile{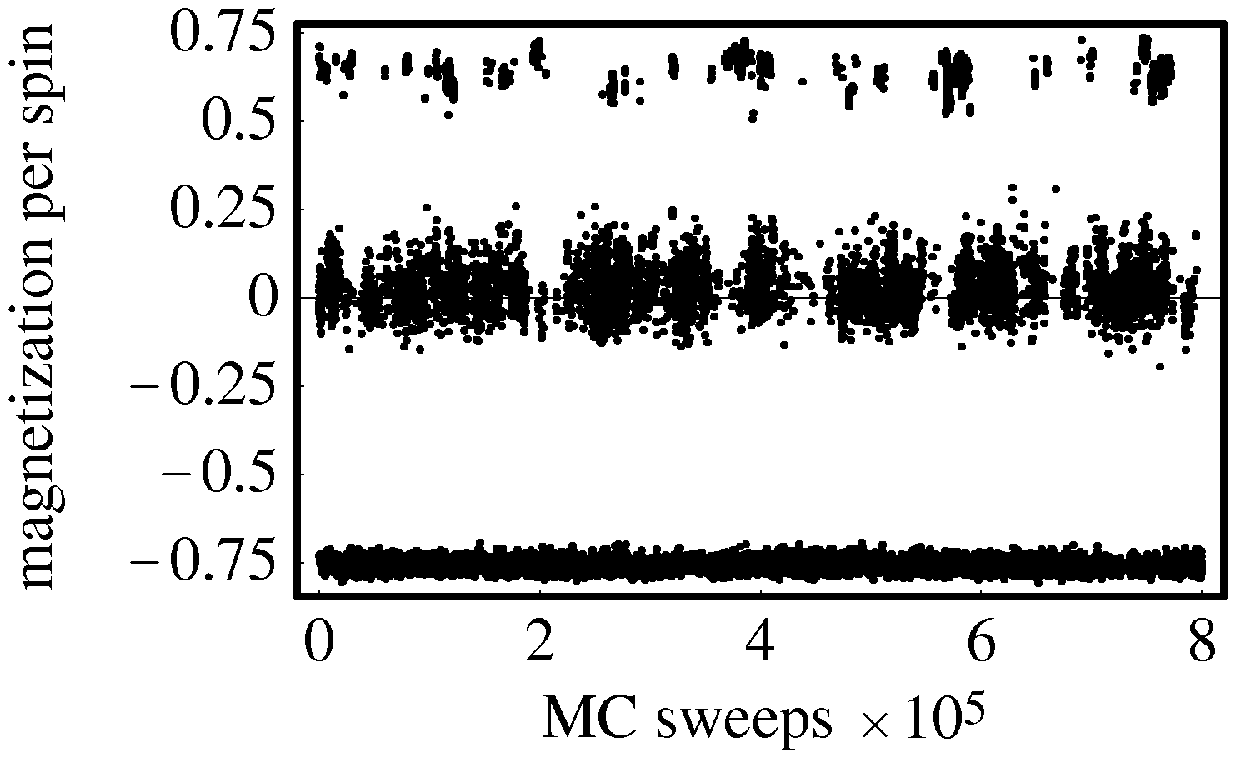}
\caption{Magnetization time series for realization 31 for $\Delta=.3267$.}
\label{fig:t31}
\end{center}
\end{figure}

We have studied the dynamics of the replica exchange algorithm and compared
it to the Metropolis algorithm.  There are two ways of comparing the
algorithms.  First, time can be measured by Monte Carlo sweeps.  This
approach favors the replica exchange algorithm since a single Monte Carlo
sweep involves flipping spins in many (here 16) replicas and since growing
clusters is computationally more intensive than flipping single spins.  The
second approach is to compare actual running time for the two algorithms on
the same computer.  This approach ignores the possibility that the one
algorithm is better optimized and it is not an appropriate way to measure
fundamental quantities such as the dynamic exponent however it does give a
practical comparison and is useful for deciding which algorithm to choose
for a given system size.  For $24^3$ systems with 16 replicas with maximum
disorder $\Delta=.35$ on a 450MHz Pentium III machine we find that the
replica exchange algorithm runs at about $0.56$ sec/sweep and that our
implementation of the Metropolis algorithm runs at about $0.0088$
sec/sweep.  As is the case for the equilibrium properties of the system, we
find great differences in the dynamics depending on the realization of
disorder.  These differences are much more pronounced for the Metropolis
algorithm.  We measured the integrated autocorrelation time for the
magnetization for realization 20 for both algorithms with the result
$\tau^{\rm Metropolis}= 6000$ and $\tau^{\rm replica}=1100$ when measured
in Monte Carlo sweeps.  The advantage of the replica exchange algorithm is
lost however when these times are converted to running times; $\tau^{\rm
  Metropolis}= 53$s and $\tau^{\rm replica}=616$s.  The joint
energy/magnetization histogram for realization 20 is smooth and without
gaps suggesting that there are no global energy barriers separating regions
of phase space.  At the other extreme is realization 31 for which the joint
energy/magnetization distribution is split into three pieces separated by
gaps where the probability is very small.  Using the parameters for 
 realization 31 determined by fine-tuning for $\Delta=.3267$, we did a 
long Metropolis run of 100 million Monte Carlo sweeps starting from random
initial spins.  The simulation stayed mainly in the ``$0$'' state with two
excursions to the ``$+$'' state.  The ``$-$'' state was never visited.  On the
other hand, two Metropolis runs of 20 million sweeps starting from all spins
up and all spins down went to the ``$0$'' and ``$-$'' states, respectively,
and stayed there for the entire simulation. Figure~\ref{fig:t31} shows the
magnetization time series from the replica exchange algorithm for
realization~31 for 800,000 sweeps.  Although it is difficult to estimate the
autocorrelation time from this series, it is clear that the simulation samples
each of the states many times.  The conclusion is that for most realizations
of disorder, at size
$24^3$ there is no great practical advantage to using the replica exchange
algorithm but that for cases where well separated regions are present in the
joint distribution, only the replica exchange algorithm is able to reach
equilibrium within reasonable simulation times.

\section{Discussion}
One of the chief motivations for this study was to determine the nature of
the RFIM phase transition.  Unfortunately, our results fail to settle this
question.  What would we have expected from the different scenarios for the
phase transition?  If the transition is first-order we should find three
phases in coexistence at the transition point.  The high temperature phase,
would have little magnetization and a large bond energy while the two low
temperature phases would have large absolute values of magnetization, one
positive and the other negative, and small bond energies.  This situation
is exactly what is seen in roughly 10$\%$ of realizations.  For systems
much smaller than $24^3$, three phase coexistence is not seen simply
because the magnetization distribution is too broad to display three
distinct peaks.  One hypothesis is that, as system size becomes large, the
fraction of systems displaying three phases increases and that, in the
thermodynamic limit, the transition at $\Delta=.35$ is first-order.
However, the existence of realization~14 with its five peaks is difficult
to reconcile with this viewpoint.

If the transition is continuous and follows the droplet model
scenario~\cite{Villain85,Fish86,EiBi} we would expect two peaks in the
magnetization histogram corresponding to two states of a single critical
phase.  The width of each peak should scale as $L^{-\gamma/\nu}$ and the
separation between the peaks should scale as $L^{-\beta/\nu}$.  The
majority of our disorder realizations display two peaks and an alternative
to the first-order transition hypothesis is that, as system size increases,
the fraction of realizations with two peaks in the magnetization histogram
approaches $100\%$.  Although the original droplet model envisioned two
states with roughly the same energy and differing by flipping a critical
cluster of spins, it is possible to imagine more complicated droplet
pictures where there are sometimes more than two states in the critical
phase and therefore more than two peaks
 in the magnetization histogram.
However, our five realizations that display more than one peak in the bond
energy histogram are difficult to reconcile with the droplet model since
one expects the states, however many there are, to be nearly degenerate in
bond energy.

Ultimately, the essential difference between the first-order scenario and
the droplet model continuous-transition scenario is whether there are
distinct phases at the transition point or whether the peaks in the
magnetization histogram are different states that are part of a single
critical phase.  The comes down to the question of whether the peaks in
magnetization histogram move toward the origin as $L\to\infty$ or remain at
finite values.  Unfortunately, the consensus is that the magnetization
exponent, if it is not actually zero, is very tiny so that the decrease in
magnetization with system size would be far too weak to observe in
simulations.

\section{Conclusions}

We have studied the phase transition of the random field Ising 
model by Monte Carlo simulation of
systems of size $24^3$ using a new cluster algorithm.  We find major
qualitative differences between the behavior of different realizations of
disorder.  Some realizations display a broad two-peaked magnetization
histogram consistent with a continuous
transition, while a small fraction display a three-peaked structure
consistent with a first-order transition.  Our main conclusion is that more
work needs to be done to determine the nature of the transition.  It would
be very useful to study larger system sizes and stronger disorder to
determine whether there are trends in the qualitative features of the phase
transition.  Does the fraction of ``first-order'' systems increase as
disorder or system size increases?  It will require improvements of the
algorithm including the fine-tuning technique or substantially more
computer power to resolve these questions.

\section{Acknowledgements}
We thank Pozen Wong and Robert Swendsen for useful discussions.  This work
was supported by NSF grants DMR-9978233 (JM) and DMS-9971016 (LC) and NSA
grant MDA904-98-1-0518 (LC).  JM thanks the Santa Fe Institute for
hospitality during a visit in which some of this work was carried out.

\epsfclipon

\begin{thebibliography}{10}

\bibitem{ImMa75}
Y.~Imry and S.~K. Ma.
\newblock Random-field instability of the ordered state of continuous symmetry.
\newblock {\em Phys. Rev. Lett.}, 35:1399, 1975.

\bibitem{Imbrie84}
J.~Z. Imbrie.
\newblock Lower critical dimension of the random-field {I}sing model.
\newblock {\em Phys. Rev. Lett.}, 53:1747, 1984.

\bibitem{BrKu87}
J.~Bricmont and A.~Kupiainen.
\newblock Lower critical dimension for the random-field {I}sing model.
\newblock {\em Phys. Rev. Lett.}, 59:1829, 1987.

\bibitem{Natt97}
T.~Nattermann.
\newblock Theory of the random field {I}sing model.
\newblock In A.~P. Young, editor, {\em Spin Glasses and Random Fields}. World
  Scientific, Singapore, 1997.

\bibitem{BrDe98}
E.~Brezin and C.~De~Dominicis.
\newblock New phenomena in the random field {I}sing model.
\newblock {\em Europhys. Lett.}, 44:13, 1998.

\bibitem{Villain85}
J.~Villain.
\newblock Equilibrium critical properties of random field systems: New
  conjectures.
\newblock {\em J. Physique}, 46:1843, 1985.

\bibitem{BrMo}
A.~J. Bray and M.~A. Moore.
\newblock Scaling theory of the random-field {I}sing model.
\newblock {\em J. Phys. C}, 18:L927, 1985.

\bibitem{Fish86}
D.~S. Fisher.
\newblock Scaling and critical slowing down in random-field {I}sing systems.
\newblock {\em Phys. Rev. Lett.}, 56:416, 1986.

\bibitem{CaMa}
M.~S. Cao and J.~Machta.
\newblock Migdal-{K}adanoff study of the random field {I}sing model.
\newblock {\em Phys. Rev. B}, 48, 1993.

\bibitem{FaBeMc}
A.~Falicov, A.~N. Berker, and S.~R. McKay.
\newblock Renormalization-group theory of the random-field {I}sing model in
  three dimensions.
\newblock {\em Phys. Rev. B}, 51:8266, 1995.

\bibitem{HoKhSe}
A.~Houghton, A.~Khurana, and F.~J. Seco.
\newblock Fluctuation-driven first-order phase transition, below four
  dimensions, in the random-field {I}sing model with a gaussian random-field
  distribution.
\newblock {\em Phys. Rev. Lett.}, 55, 1985.

\bibitem{GoAdAhHaSc}
M.~Gofman, J.~Adler, A.~Aharony, A.~B. Harris, and M.~Schwartz.
\newblock Critical behavior of the random field {I}sing model.
\newblock {\em Phys. Rev. B}, 54:364, 1996.

\bibitem{RiYo}
H.~Rieger and A.~P. Young.
\newblock Critical behavior of the three-dimensional random-field {I}sing
  model: Two-exponent scaling and discontinuous transition.
\newblock {\em Phys. Rev. B}, 52:6659, 1995.

\bibitem{Rieg95}
H.~Rieger.
\newblock Critical behavior of the three dimensional random field {I}sing
  model.
\newblock {\em J. Phys. A: Math. Gen.}, 26:5279, 1993.

\bibitem{NeBa}
M.~E.~J. Newman and G.~T. Barkema.
\newblock {M}onte {C}arlo study of the random-field {I}sing model.
\newblock {\em Phys. Rev. E}, 53:393, 1996.

\bibitem{YoNa}
A.~P. Young and M.~Nauenberg.
\newblock Quasicritical behavior and first-order transition in the $d=3$
  random-field {I}sing model.
\newblock {\em Phys. Rev. Lett.}, 54:2429, 1985.

\bibitem{Ogielski}
Ogielski.
\newblock Integer optimization and zero-temperature fixed point in {I}sing
  random-field systems.
\newblock {\em Phys. Rev. Lett.}, 57:1251, 1986.

\bibitem{Sour98}
N.~Sourlas.
\newblock Universality in random systems: The case of the 3-d random field
  {I}sing model.
\newblock {\em Comput. Phys. Commun.}, 122:183, 1999.

\bibitem{SwWa86}
R.~H. Swendsen and J.-S. Wang.
\newblock Replica {M}onte {C}arlo simulations of spin glasses.
\newblock {\em Phys. Rev. Lett.}, 57:2607--2609, 1986.

\bibitem{ReMaCh}
O.~Redner, J.~Machta, and L.~F. Chayes.
\newblock Graphical representations and cluster algorithms for critical points
  with fields.
\newblock {\em Phys. Rev. E}, 58:2749--2752, 1998.

\bibitem{ChMaRe98b}
L.~Chayes, J.~Machta, and O.~Redner.
\newblock Graphical representations for {I}sing systems in external fields.
\newblock {\em J. Stat. Phys.}, 93:17, 1998.

\bibitem{MaPa}
E.~Marinari and G.~Parisi.
\newblock Simulated tempering: A new {M}onte {C}arlo scheme.
\newblock {\em Europhys. Lett.}, 19:451, 1992.

\bibitem{Wolff}
U.~Wolff.
\newblock Collective {M}onte {C}arlo updating for spin systems.
\newblock {\em Phys. Rev. Lett.}, 62:361, 1989.

\bibitem{NeBa99}
M.~E.~J. Newman and G.~T. Barkema.
\newblock {\em Monte Carlo Methods in Statistical Physics}.
\newblock Oxford, Oxford, 1999.

\bibitem{EiBi}
K.~Eichorn and K.~Binder.
\newblock Monte carlo investigation of the three-dimensional random-field
  three-state {P}otts model.
\newblock {\em J. Phys. A: Condens. Matter}, 8:5209--5227, 1996.

\end{thebibliography}
\end{document}